\documentclass[%
preprint,
 amsmath,amssymb,
 aps,
]{revtex4-1}

\usepackage{tikz}
\usepackage{graphicx}
\usepackage{dcolumn}
\usepackage{bm}
\usepackage{hyperref}

\begin{document}


\title{A Factorisation Algorithm in\\ Adiabatic Quantum Computation}

\author{Tien D. Kieu}
 \email{tien.d.kieu@gmail.com}
 \email{tdkieu@swin.edu.au}
\affiliation{%
Centre for Quantum and Optical Science\\
Swinburne University of Technology, Victoria, Australia
}%




\date{\today}

\begin{abstract}
The problem of factorising positive integer $N$ into two integer factors $x$ and $y$ is first reformulated as an optimisation problem over the positive integer domain of either of the Diophantine polynomials $Q_N(x,y)=N^2(N-xy)^2 + x(x-y)^2$ or $R_N(x,y) = N^2(N-xy)^2 + (x-y)^2 + x$, of each of which the optimal solution is unique with $x\le \sqrt{N} \le y$, and $x=1$ if and only if $N$ is prime. An algorithm in the context of Adiabatic Quantum Computation is then proposed for the general factorisation problem.

\end{abstract}

\pacs{Valid PACS appear here}
    \maketitle


Factoring an integer into its prime constituents has attracted much interest since the advance of the RSA public-private key encryption~\cite{RSA}.  It is suspected that factorisation is NP-intermediate, that is, in the NP class but may be not quite NP-complete.  While there does not yet exist any polynomial-time algorithm for the problem on a classical/Turing computer, the discovery of Shor's quantum algorithm with quantum circuits~\cite{Shor} has been one of the main motivations for research into quantum computation and building of quantum computers.

In this paper we also consider the factorisation problem in the realm of quantum computation but with adiabatic processes, in complementary addition to the computation with quantum circuits.  Also recently, the authors of~\cite{Jiang} have considered the problem with quantum annealing.  We first reformulate in the next section  the factorisation into two integer factors as an optimisation of some corresponding Diophantine polynomials over the integer domain.  The optimisation could also be repeatedly applied to any integer having more than two prime factors. Based on this reformulation, we then present an algorithm in the context of AQC (Adiabatic Quantum Computation) for the general factorisation problem.  Following that are some numerical illustrations of the algorithm and discussion on the lower bound of the computing time with the help of an energy-time uncertainty relation. The paper is then concluded with some remarks.

\section*{Factorisation as an optimisation problem}
We first consider the problem of factorising a natural integer $N$ into two integer factors $x$ and $y$.  We propose that this problem can be reformulated as an optimisation problem over the integer domain of the following Diophantine polynomial
\begin{eqnarray}
\min_{x, y\in \mathbb{N}} \left\{Q_N(x,y) \equiv N^2(N-xy)^2 + x(x-y)^2 \right\}.  \label{polynomial}
\end{eqnarray}
Without the second term in $Q_N$ above, the optimal solutions contain the trivial unity factor. We could eliminate this triviality with this second term replaced by $(x-y)^2$, but there still remains a symmetry between $x$ and $y$.
We now show that the polynomial $Q_N(x,y)$ in~(\ref{polynomial}) is minimised if and only if $xy=N$ where $x\le y$ and $x$ is nearest to, but not exceeding, the integer part of $\sqrt{N}$.  The optimal solution thus has no trivial factor $x=1$ unless $N$ is prime.

For $xy=N$ the first term on the rhs of~(\ref{polynomial}) vanishes and the remaining second term is obviously smaller for $1 \le x\le \sqrt{N} \le y$.  In this case, the second term is a one-variable function $x(x-N/x)^2$, and it is not difficult to see that this term is a decreasing function for $1 \le x \le \sqrt{N}$.  We then have
\begin{eqnarray}
\left. Q_N(x,y)\right|_{N=xy} = Q_N(x,1/x) &\le& \max_{1 \le x \le \sqrt{N}}{x(x-N/x)^2}, \nonumber\\
&\le& \left.x(x-N/x)^2\right|_{x=1}, \nonumber\\
&\le& (N-1)^2. \label{xy_equal_y}
\end{eqnarray}
The second line results from the fact that $x(x-N/x)^2$ obtains its maximum value at the boundary where $x=1$.  The closer $x$ is to the integer part of $\sqrt{N}$ the smaller the value of $x(x-N/x)^2$.  

Now if $xy\not=N$ then obviously $(N-xy)^2 \ge 1$ and the first term of~(\ref{polynomial}) is consequently not less than $N^2$.  Thus
\begin{eqnarray}
\left.Q_N(x,y)\right|_{N\not=xy} &\geq& N^2 + x(x-y)^2, \nonumber\\
&\geq& N^2. \label{xy_not_N}
\end{eqnarray}
Combining~(\ref{xy_equal_y}) and~(\ref{xy_not_N}), we have for $N>1$
\begin{eqnarray}
\left.Q_N(x,y)\right|_{N\not=xy} &>& \left.Q_N(x,y)\right|_{N=xy}. \label{hierachy}
\end{eqnarray}

The second term $x(x-y)^2$ of $Q_N(x,y)$ in~(\ref{polynomial}) is thus designed to introduce an asymmetry between $x$ and $y$, enforcing $x\le y$ at the optimum value, and to eliminate the trivial unity factor of $N$ unless $N$ is prime.  The optimisation problem~(\ref{polynomial}) thus has a {\it unique} solution $(x,y)$ for $xy=N$ with $x\le \sqrt{N} \le y$.  In general, $x$ is the integer factor nearest to, but not exceeding, $\sqrt{N}$.  Consequently, $x=1$ if and only if $N$ is prime. 

The optimisation of $Q_N(x,y)$ can thus also determine the primality of $N$.  

For $N$ having more than one pair of two prime factors, the optimisation of $Q_N(x,y)$ still admits only one {\it unique} solution to yield a non-trivial factoring pair of $N$.  For $N$ having more than two prime factors we can apply the optimisation repeatedly for the successive factors to obtain the constituent prime factors.

\begin{figure}
\begin{center}
\includegraphics[scale=0.7]{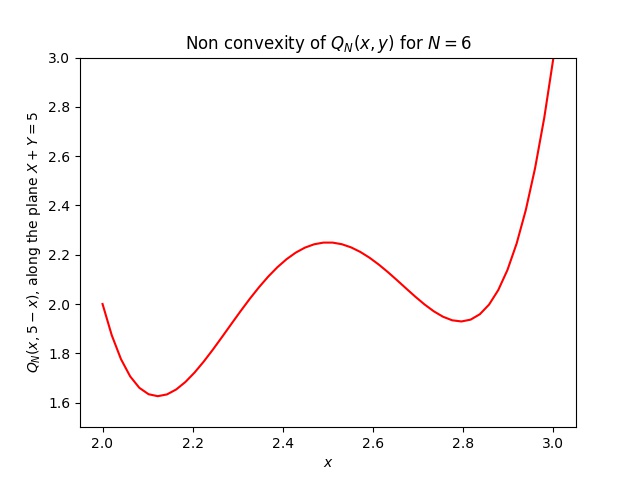}
\caption{Illustration of non-convexity of $Q_N(x,y)$ along the plane $x+y=5$ for $N=6$.}
\label{fig:figure1}
\end{center}
\end{figure}

It can be shown that the objective function in~(\ref{polynomial}) is non-convex as depicted in FIG.~\ref{fig:figure1} for a slice of the polynomial for $N=6$ along the plane $y+x=5$. (In general, determining convexity for multivariate quartic polynomials is NP-hard.)

It is also known that the problem of minimising a general non-convex degree four polynomial over a two-dimensional convex polygon like that in~(\ref{polynomial}) is classically NP-hard -- even though there exist polynomial-time approximations for some lower and upper bounds of the optimal values; see, for example,~\cite{Loera} and references therein.  The condition of convexity or lack of it is an important element here; were $Q_N(x,y)$ convex then we would already have a polynomial time algorithm for the factorisation!

  In the next section we propose an AQC algorthim for optimising this type of polynomials over the integer domain.

\section*{An AQC algorithm for the factorisation problem}
AQC~\cite{Farhia,Farhib} makes use of some appropriate time-varying Hamiltonians and an initial state.  The computation starts with the readily constructible ground state $|g_I\rangle$
of an initial Hamiltonian $H_I$ which is then {\em adiabatically}
extrapolated to the final target Hamiltonian $H_P$ whose ground state
$|g\rangle$ encodes the desirable
solution of the problem and could be then obtained with
reasonably high probability.	The interpolation between $H_I$ and
$H_P$ is facilitated by a time-dependent Hamiltonian in the time
interval $0\le t \le T$,
\begin{eqnarray}
{\cal H}(t) &=& f(t/T)H_I + g(t/T)H_P,
\label{extrapolation}
\end{eqnarray}
either in a temporally linear manner (that is, $f(t/T) =
\left(1-{t}/{T}\right)$ and $g(t/T)={t}/{T} $); or otherwise with $f(0) = 1 = g(1)$ and $f(1) = 0 = g(0)$.  We also assume that both $f$ and $g$ are continuous, and $g$ is semi-positive for all $t\in[0,T]$.	Such a time
evolution is captured by the following Schr\"odinger equation in the time interval $[0,T]$
\begin{eqnarray}
i\partial_t |\psi(t)\rangle &=& {\cal H}(t)\;|\psi(t)\rangle,
\label{AQC}\\
|\psi(0)\rangle &=& |g_I\rangle. \nonumber
\end{eqnarray}
As the rate of the evolution of the Hamiltonian approaches zero, the end state $|\psi(T)\rangle$ asymptotically converges to the target state $|g\rangle$ as asserted by the quantum adiabatic theorem~\cite{Messiah}.

\subsection*{The initial Hamiltonian $H_I$ and its ground state $|g_I\rangle$}
Corresponding to the variable $x$, we introduce the operators $a^\dagger_{x}$ and $a_{x}$,
which respectively create and annihilate the number states $|n_x\rangle$, for $n_x=0,1, \ldots$:
\begin{eqnarray}
\left[a_{x}, a^\dagger_{x}\right] &=& \bf{1}, \label{commR}\\ 
a^\dagger_{x} |n_{x}\rangle &=& \sqrt{n_x+1}|n_{x}+1\rangle, \nonumber\\
a_{x} |n_{x}\rangle &=& \sqrt{n_x}|n_{x}-1\rangle, n_x\ge 1,\nonumber\\
a_{x} |0_x\rangle &=& 0,  \nonumber
\end{eqnarray}
and the number operators $\hat{n}_{x}$ are constructed in the usual way:
\begin{eqnarray}
\hat{n}_{x} &=& a^\dagger_{x}a_{x}, \nonumber\\
\hat{n}_{x}|n_x\rangle &=& n_x|n_x\rangle.
\end{eqnarray}
Similarly for the variable $y$ with $a^\dagger_{y}$, $a_{y}$ and $\hat{n}_y$.  

We propose to start the AQC with the following initial Hamiltonian $H_I$:
\begin{eqnarray}
 H_I &=& (a^\dagger_{x} - \theta_{x}^*)(a_{x} - \theta_{x}) + (a^\dagger_{y} - \theta_{y}^*)(a_{y} - \theta_{y}), \label{H_I}
\end{eqnarray}
for some c-numbers $\theta_{x}$ and $\theta_y$.  The states in the total Hilbert space can be  now decomposed in terms of $|n_x\rangle\otimes|m_y\rangle$, the direct products of the two bases of number states.  This Hamiltonian admits the readily constructible direct product of coherent states as the ground state $|g_I\rangle$,
\begin{eqnarray}
|g_I\rangle &=& |\theta_x\rangle \otimes |\theta_y\rangle, {\rm \;shortened \; as \;} |\theta_x \theta_y\rangle,
\end{eqnarray}
with $|\theta\rangle$ the canonical quantum coherent state
\begin{eqnarray}
|\theta\rangle &=& {\rm e}^{-{\frac{|\theta|^2}{2}}}\sum_{n=0}^\infty \frac{\theta^{n}}{\sqrt[]{n!}}|n\rangle, \nonumber
\end{eqnarray}
where:
\begin{eqnarray} 
a\; |\theta\rangle &=& \theta\;|\theta\rangle. \nonumber
\end{eqnarray}

\subsection*{The target Hamiltonian $H_P$}
Building on~(\ref{polynomial}), we introduce the target Hamiltonian $H_P$,
\begin{eqnarray}
H_P &=& N^2(N-\hat{n}_x\hat{n}_y)^2 + \hat{n}_x(\hat{n}_x-\hat{n}_y)^2, \label{H_P}
\end{eqnarray}
which has a unique ground state, i.e. no degeneracy.  This ground state would give us the solution for the desirable optimisation problem~(\ref{polynomial}) over the integer domain, thus solving the factorisation for the integer $N$.

Our proposed AQC may be physically realised by quantum optics or other means, with some suitably chosen end time $T$.  Then the solution factors for $N$ would be obtained from the final state $|\psi(T)\rangle$ by its measurements in the number state basis.  Even though such measurements only yield probabilities, any factor candidates so obtained could be verified straightforwardly and efficiently by a multiplication of $n_x$ and $n_y$, or by a division of $N$ by $n_x$ or by $n_y$.

After the release of a draft of this present work, the algorithm presented herein has been physically implemented with qubits on D-Wave Systems~\cite{Warren-Dahl}.

\section*{Numerical simulations}
\begin{figure}
\begin{center}
\includegraphics[scale=0.7]{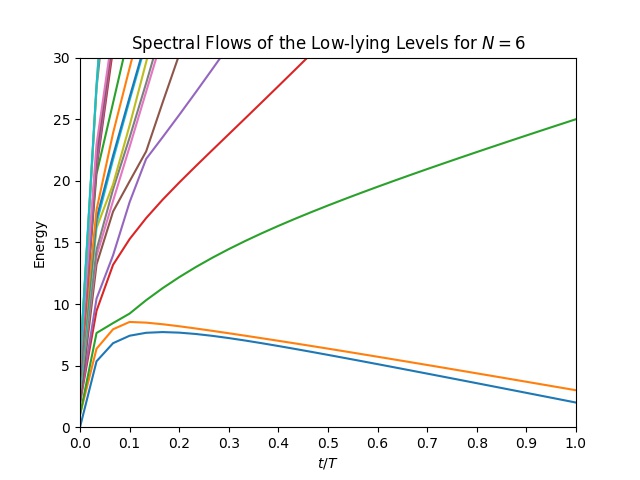}
\caption{Spectral flow of ${\cal H}(t)$ with linear extrapolation in~(\ref{extrapolation}) for $N=6$.  The two lowest branches respectively correspond  at $t=T$ to $|2_x3_y\rangle$ and $|3_x2_y\rangle$, both of which are non-trivial factor pairs.}
\label{fig:figure3}
\end{center}
\end{figure}

\begin{figure}
\begin{center}
\includegraphics[scale=0.7]{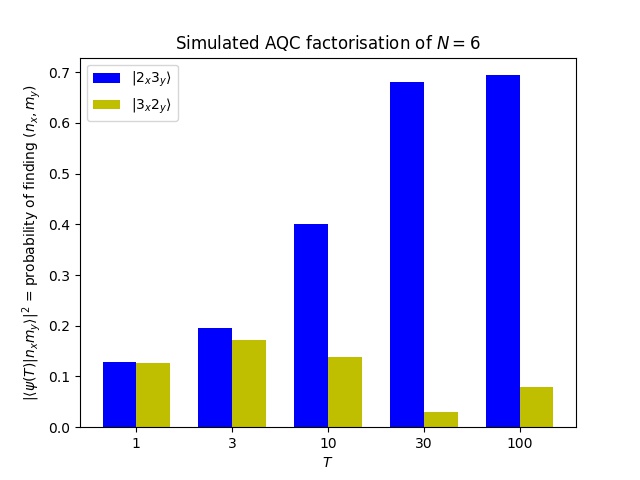}
\caption{Probability $|\langle\psi(T)|n_x m_y\rangle|^2$ versus $T$ in the simulated AQC for the factorisation of $N=6$ with a suitably truncated Hilbert space of dimensions $23\times 23$.  Only the two largest probabilities are shown.}
\label{fig:figure2}
\end{center}
\end{figure}

For the purpose of illustration, we simulate the AQC algorithm for $N=6$ by numerically solving the Schr\"odinger equation~(\ref{AQC}) in a suitably truncated Hilbert space of dimensions $23\times 23$, with linear extrapolation in~(\ref{extrapolation}) and parameters $\theta_x=\theta_y=\sqrt[4]{N}$ (so that $\langle \theta_x \theta_y | \hat{n}_x\hat{n}_y |\theta_x \theta_y \rangle = N$).  

FIG.~\ref{fig:figure3} depicts the spectral flow of ${\cal H}(t)$.  The two lowest branches respectively correspond  at $t=T$ to $|2_x3_y\rangle$ and $|3_x2_y\rangle$, both of which are non-trivial factor pairs of $N=6$.  Note the relatively wide energy gaps between these solution states and the non-solution $n_xm_y\not=N$.  This would be helpful in shortening the running time with an appropriately chosen $g(t/T)$ for the AQC algorithm, in accordance with~(\ref{time1}) in the next Section.

FIG.~\ref{fig:figure2} shows the probabilities $|\langle\psi(T)|n_x m_y\rangle|^2$ versus $T$ for the two largest probabilities.  In all instances of $T$ there, the number state $|2_x3_y\rangle$ has the maximum probability among all the number states.  This probability increases with $T$, the inverse of the evolution rate, in agreement with the quantum adiabatic theorem~\cite{Messiah}.  The next highest probability corresponds to the other solution $|3_x2_y\rangle$.

We note that the increased probability for $|3_x 2_y\rangle$ at $T=100$ compared to that at $T=30$ is an artifact of our truncation of the Hilbert space.  With a truncation at $|n_{\rm max}\rangle$ such that $a^\dagger |n_{\rm max}\rangle = 0$,
the commutation relation~(\ref{commR}) is violated
\begin{eqnarray}
\langle n_{\rm max}|aa^\dagger |n_{\rm max}\rangle &=& 0, \nonumber\\
\langle n_{\rm max}|a^\dagger a + {\bf 1} |n_{\rm max}\rangle&=& n_{\rm max} + 1.
\end{eqnarray}
The effect of such a truncation artifact becomes more prominent for larger $T$, unless we increase $n_{\rm max}$ accordingly.

\section*{Computational complexity}
We have presented elsewhere~\cite{Kieu2} a necessary condition for a lower time limit required in a general AQC for an initial state to  evolve into an orthogonal state under the dynamics of ${\cal H}$.  Namely, it is necessary that the evolution time cannot be less than $T_\perp$,
\begin{eqnarray}
T_\perp &\sim& {\cal O} \left(\frac{1}{\Delta_I E_P\int_0^1 g(\tau) d\tau}\right), \label{time1}
\end{eqnarray}
where $\Delta_I E_P$ is the energy spread of the initial state $|g_I\rangle$ in terms of the target Hamiltonian $H_P$,
\begin{eqnarray}
\Delta_I E_P &\equiv& \sqrt{\langle g_I| H_P^2 |g_I\rangle - \langle g_I| H_P |g_I\rangle^2}.
\end{eqnarray}
It is important to note that only the {\em initial} eigenstate $|g_I\rangle$ (which is of course {\em time-independent}), and neither the instantaneous eigenstates nor the full time-dependent wave function at any other times, is required for the time condition~(\ref{time1}).  This hallmark of our results in~\cite{Kieu2} enables their wider applicability and usefulness.

We note that $T_\perp$ is particularly a function of the parameters $|\theta|$,
\begin{eqnarray}
T_\perp &=& T_\perp(|\theta|).
\end{eqnarray}
The parameters $\theta$'s give our proposed AQC algorithm some advantage that is denied or not evident elsewhere,
and this could be exploited in order to {\rm reduce} the lower time limit $T_\perp$ for the computation to be ${\cal O}(1)$ or even less!  But this reduction would at the same time incur an appropriate increase in the energy cost $E_{\rm cost}=\max_{0\le t \le T} \langle g_I|{\cal H}(t)|g_I\rangle$.  For instance, for $\theta \sim {\cal O}(1)$ the cost $E_{\rm cost}$ would be, as evident from~(\ref{H_P}), of order ${\cal O}(N^4)$ -- that is, exponentially in the number of bits in the binary representation of $N$.

It is not that surprising that we could reduce the computation time with more energy resources.  Other examples of this delicate balance between the energy required and the lower time limit for a well known quantum algorithm are given elsewhere~\cite{Kieu2, KieuTSP}.  



\section*{Another Diophantine polynomial}
We could also employ another Diophantine polynomial $R_N(x,y)$, in addition to $Q_N(x,y)$ of~(\ref{polynomial}), to construct $H_P$ in the search to find a pair of factors for $N$,
\begin{eqnarray}
\min_{x, y\in \mathbb{N}} \left\{R_N(x,y) \equiv N^2(N-xy)^2 + x(x-y) + x \right\}.  \label{polynomial2}
\end{eqnarray}
The proof that the optimisation of $R_N(x,y)$ could yield a non-trivial pair of factors is similar to that for~(\ref{polynomial}).

For $xy\not=N$ then obviously $(N-xy)^2 \ge 1$ and the first term of~(\ref{polynomial2}) is consequently not less than $N^2$,
\begin{eqnarray}
\left. R_N(x,y)\right|_{N\not=xy} &\ge& N^2.
\end{eqnarray}

For $xy=N$ the first term on the rhs of~(\ref{polynomial2}) vanishes and the remaining terms, for $x \le \sqrt{N} \le y$,
\begin{eqnarray}
\left. R_N(x,y)\right|_{N=xy} = R_N(x,1/x) &=& \left(x - \frac{N}{x}\right)^2 + x
\end{eqnarray}
form a convex function (i.e. having non-negative second derivative) for $x \in [1,\sqrt{N}]$.  It then follows that such a function is bounded by its maximum value obtained at one of the end points
\begin{eqnarray}
\left.R_N(x,y)\right|_{N=xy} = R_N(x,1/x) &\le& \max\left\{\left.R_N(x,1/x)\right|_{x=1}, \left.R_N(x,1/x)\right|_{x=\sqrt{N}}\right\},\nonumber\\
&\le& \max\left\{(1-N)^2+1, \sqrt{N}\right\},\label{19}\\
&\le& (1-N)^2+1, \nonumber\\
&\le& N^2,\nonumber\\
&\le& \left.R_N(x,y)\right|_{N\not=xy}. \nonumber
\end{eqnarray}

A pair of factors of $N$ so obtained is not only unique but also {\it non-trivial}, because from~(\ref{19}) it follows that $x>1$ when $N$ is not prime.  Only when $N$ is prime then we would have $x=1$.

\section*{Concluding Remarks}
We have mapped the factorisation problem into the optimisation over the integer domain of some corresponding Diophantine polynomials.  The optimal solutions are unique, and contain the unity factor if and only if the integers to be factored are prime.  Based on this, we then propose an algorithm in the context of Adiabatic Quantum Computation for the general factorisation.  We 
also indicate that the lower bound on the running time could be reduced
at the expense of more energy input in order to carry out the physical computation.  Some numerical simulations of the algorithm for a simple case are also presented for illustration purpose.

After the release of a draft of this present work, the algorithm presented herein has been physically implemented with qubits on D-Wave Systems~\cite{Warren-Dahl}.

We are grateful for discussions with Peter Hannaford, Adolfo del Campo and Richard Warren, who also brought the work~\cite{Jiang} to our attention after the completion of this paper.  The publication fee for this work is covered by Swinburne University.

\bibliography{Factorisation}

\end{document}